\documentclass{aa}             
\usepackage{graphicx}
\usepackage{txfonts}
\begin{document}

\title{Why soft X-ray transients can remain in the low/hard state 
during outburst}
\author{ E. Meyer-Hofmeister}
\offprints{Emmi Meyer-Hofmeister}
\institute{Max-Planck-Institut f\"ur Astrophysik, Karl-
Schwarzschildstr.~1, D-85740 Garching, Germany
} 

\date{Received: / Accepted:}

\abstract{
In the canonical understanding of transient X-ray sources the accretion 
during quiescence occurs via a geometrically thin disk in the outer part
and via an advection-dominated hot coronal flow/ADAF in the inner part.
The inner part important for the radiation yields a hard spectrum. 
In most sources the luminosity increase during outburst causes a change over
to a soft spectrum which can be understood as that of a multi-color
black body disk reaching inward to the marginally stable orbit. A few 
transient sources do not display this transition to the soft state. We
show that this can be understood as due to relatively low peak
luminosities in outburst in systems with short orbital periods and
therefore less mass accumulated in the then smaller accretion disk. 
This is in agreement with the observations which show that most likely 
short orbital period systems remain in the hard spectral state.

\keywords{Accretion, accretion disks -- black hole physics  --
binaries, close -- stars: individual: XTE J1118+480, GRO J0422}
}
\titlerunning {}
\maketitle

\section{Introduction}
Accretion onto stellar-mass black holes in binaries is of interest in
connection also with supermassive black holes with
relevance to galaxy formation and cosmological questions. 
Since the first identification of a black hole X-ray binary, Cygnus
X-1, by Webster and Murdin (1972) and Bolton (1972), 30 years ago, several 
black hole X-ray binaries, persistently bright or transient were
found. According to the nature of the secondary star systems with a high-mass 
companion, usually a massive O/B star, are classified as high-mass
X-ray binaries (HMXB) and systems with a low-mass companion, of a few 
$M_\odot$ or less, as low-mass X-ray binaries (LMXB). 18 confirmed
black hole binaries are now identified. A description of their
properties and a discussion of the relevant issues is given in the
recent review by McClintock \& Remillard (2003), a description of the 
observations in different wavelength regions can be found in the review by
Charles and Coe (2003). The gas flowing towards the black
hole originates either from a wind of the high-mass companion (in HMXBs)
or from Roche lobe overflow (in LMXBs). In both cases an accretion disk
forms around the compact star.

Well known are the wind accreting systems with black holes of 
$\ge 10 M_\odot$ and high accretion rates, accordingly a fully
ionized accretion disk and appearing as persistently bright X-ray
sources. Based on a recent high-resolution survey carried out with 
Chandra, Wang et al. (2002) report on faint sources which might also be wind 
accreting sources, but mainly neutron star wind accretors (Pfahl et
al. 2002).
Mass overflow rates from the low-mass companion stars are usually
lower so that the gas is accumulated in a cool quiescent disk
until an outburst is triggered. These outburst cycles caused by a disk 
instability are a feature which LMXBs  have in common with dwarf novae.
In several LMXBs the outburst recurrence time is so long that
only one outburst was observed since the beginning of X-ray
observations. The systems only detected during such an outburst
were called X-ray novae. During quiescence the accretion rate is low,
an advection-dominated accretion region fills the inner part,
and the spectrum is hard. Usually in outburst the spectrum changes to
a soft multi-color black body spectrum, which is interpreted as the
radiation from a geometrically thin disk reaching inward to the
innermost stable orbit. Such changes from a low/hard to a high/soft
state and back are observed for several black hole LMXB outbursts and 
also in persistently bright HMXBs as well as in neutron stars.

Surprisingly a few LMXBs remain in the low/hard state during outburst, 
best observed for XTE J1118+480 and GRO J0422+32. It turns out that whether
a transition to the high/soft state occurs or not depends on the peak mass
flow rate in outburst. Only if this mass flow is high enough to overcome
evaporation the disk can reach to the innermost stable orbit 
and the spectrum can become soft. The process of evaporation of matter
from the disk into a corona therefore
is a key feature for the understanding of the spectral transitions.
The maximal evaporation rate, critical for spectral state transition, 
was determined from the coronal evaporation model (Meyer et al. 2000a)
Thus for low peak luminosity the X-ray binary sources might
remain in the low/hard state. The aim of the present investigation is 
to inquire about this special case studying the matter
accumulation in the thin disk in dependence on the parameters disk
size (orbital period), black hole mass, outburst recurrence time
and the resulting peak luminosity. It is an interesting check of the 
evaporation model to compare the luminosity at which the state
transition is observed (Maccarone 2003) with the theoretical 
prediction.

In Sect.2 we briefly review the basic modes of accretion and the
change over from disk accretion to an ADAF. In Sect. 3 we describe the
dependence of the accumulation of gas in the accretion disk during quiescence
on the system parameters. The resulting expected peak luminosity
reached in outburst and the consequences for the spectral transition
are discussed in Sect.4. In Sect.5 we compare with observations.
Discussion and conclusions follow.

\section{Accretion disks in X-ray binaries}
\subsection{Disk instability model}
As found in connection with the understanding of the cause of dwarf
nova outbursts (Osaki 1974, Meyer \& Meyer-Hofmeister 1981, Smak 1984),
for a certain range of surface density,
the disk can be either in a cool state, where hydrogen and helium are 
unionized and the mass flow rate is low or in a hot state where the
gas is ionized and the mass flow rate is high. In the cool state mass
is accumulated until a surface density is reached where only the hot
state is possible. The visual luminosity can then increase by 
more than a factor 100. It is generally accepted that the outbursts of
X-ray novae are caused by the same mechanism.

But some additional features are important.
The orbital periods of black hole X-ray binaries are longer than those
of dwarf novae because the companion stars have larger Roche lobes due to 
higher mass. Irradiation affects the disk in X-ray binary outbursts
(Mineshige \& Kusunose 1993) and also can lead to accretion of most of
the disk matter onto the black hole in the late outburst 
(King \& Ritter 1998).
An even more important feature for the disk evolution in quiescence
and the outburst is the evaporation of the inner disk, that causes the 
change over from accretion via a thin disk to a hot coronal flow 
(Meyer et al. 2000b).  Due to this evaporation process a continuous
mass flow towards the black hole prolongs the time needed for the 
accumulation of matter to a
critical surface density. For a system like A0620-00 only
$\frac{1}{3}$ of the gas transferred from the companion star is
accumulated in the disk (Meyer-Hofmeister and Meyer 1999). This
explains the very long outburst repetition time. Since the evaporation
efficiency increases strongly with the mass
of the compact star evaporation is much more important in accretion
disks around black holes than in dwarf nova accretion disks around 
white dwarfs (Meyer-Hofmeister 1998, Mineshige et al. 1998).

\begin{figure}[ht]
\begin{center}
\includegraphics[width=8.5cm]{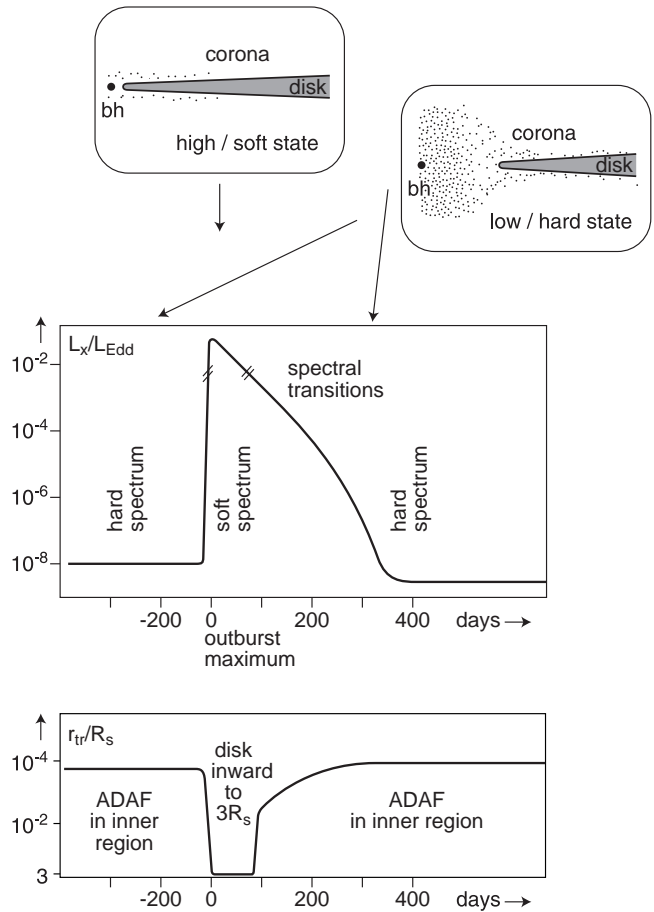}
\caption{Schematic description of an X-ray nova outburst: X-ray
luminosity $L_{\rm X}$ measured in Eddington luminosity $L_{\rm{Edd}}$,
transition radius $r_{\rm{tr}}$ from a thin disk to a pure coronal flow
measured in Schwarzschild radius $R_{\rm S}$, and related spectral 
states (numbers appropriate for A0620-00)}.
\end{center}
\end{figure}

\subsection{Critical accretion rate for the spectral state transition}
The rate at which gas evaporates from the thin cool disk into the
corona increases with decreasing distance from
the black hole. An inner edge of the disk results where
all the mass flow is transferred to the corona, inside accretion then
occurs via a pure coronal flow/ADAF. Thus the evaporation efficiency 
(Meyer et al. 2000b) determines a disk truncation. This efficiency
however reaches a maximum at a distance of about 300 Schwarzschild
radii. The consequence is that for mass flow rates higher than this maximum
the disk cannot be truncated anymore, instead continues inward to the 
last stable orbit. The critical mass flow rate corresponds to 0.02 
$L_{\rm x}/L_{\rm{Edd}}$ ($L_{\rm{Edd}}=1.3\cdot10^{38}(M/M_\odot)\ \rm{erg
s^{-1}}$).  Taking into account some uncertainty in 
Fig. 1 we mark a range of 0.02 to 0.05 $L_{\rm x}/L_{\rm{Edd}}$ (gray area).

\subsection{Advection-dominated accretion}
The X-ray flux observed from the X-ray nova A0620-00 in quiescence  
is very low compared to the optical flux from the outer disk regions
(McClintock et al. 1995). This low flux can be explained if most of the
thermal and kinetic energy is carried into the black hole by an
advection-dominated accretion flow (ADAF) and is not radiated away.
The successful description of the observed spectra (Narayan et
al. 1997, 1999) confirmed this picture. Generally the 
two-temperature ADAF model provides a natural explanation for the low 
luminosities (see review Narayan, Garcia \& Mc Clintock 2002).
The coronal gas in the inner disk region emits a hard spectrum. In
outburst when the mass flow rate increases above the critical
rate the ADAF region disappears and the spectrum changes to 
a soft multi-color black body spectrum (Esin et al. 1998). These
changes are observed for several LMXB outbursts and also in persistently 
bright HMXBs. Esin et al. (1997)  
described the configuration of the accretion flow in different states
as the very high, high, intermediate, low and quiescent state. Our
investigation concerns the high and low state. The quiescent state is
not distinct from the low state, is just a version of extremely low mass 
flow (Narayan et al. 2002).

Fig. 1 gives a schematic description of the changes of the luminosity, 
the location of the inner edge of the disk and the spectral states
during an X-ray nova outburst.

In the following investigation we use ``advection-dominated
accretion flow'' as the alternative to 
thin disk accretion, regardless of the
specific forms and modifications suggested by various authors to
meet the constraints from the observations in different wavelengths.
Since the physics of the hot accretion flow seems to be very complex
we limit our analysis to the investigation of the spectral state
transition as triggered by the balance between evaporation efficiency 
and mass flow rate in the accretion disk.

\section{Accumulation of gas in the accretion disk during quiescence}
The amount of mass accumulated in the accretion disk during quiescence
determines the peak mass flow rate in the outburst and therefore whether the 
hole caused by evaporation in the inner disk disappears.

In earlier work  (Meyer-Hofmeister \& Meyer 1999, 2000)
we investigated the mass accumulation
in disks of classical X-ray novae for a variety of system parameters,
especially appropriate for modeling A0620-00.

\begin{figure}[hb]
\includegraphics[width=6.5cm]{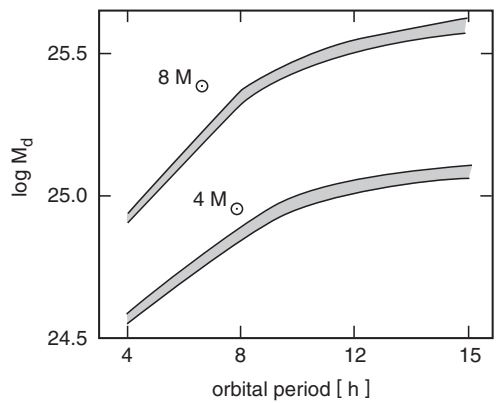}
\caption
{Mass
 $M_{\rm d}$[g] accumulated in the accretion disk at
the onset of the X-ray nova outburst. The range of values for given
black hole mass and orbital period results from
different mass overflow rates from the companion star (
corresponding to different recurrence times. We consider the situation
for matter accumulation in disks for recurrence times
$\ge$ 10 years) (from Meyer-Hofmeister \& Meyer 2000, Fig. 3).
}
\end{figure}

The accumulation of gas in the disk depends
on the mass overflow rate from the secondary star, on the disk size, 
and on the mass of the black hole. The black hole mass is an important 
parameter,
a larger mass causes a more extended hole in the inner disk and the
critical surface density can only be reached at a radius larger than
that in the case of a lower mass primary with less evaporation.
The amount of mass accumulated increases with disk size, i.e. 
orbital period. The mass overflow rate has a small effect only. 
A higher rate leads to an earlier outburst. The range of values we show 
corresponds to a recurrence time $\ge$ 10 years. In Fig. 2 we show the
amount of gas accumulated at outburst onset as a function of the
orbital period for two black hole masses. The values depend on the
frictional parameter $\alpha$. We modeled A0620-00 using the value 
$\alpha$=0.05. We were able to find agreement in several aspects:
(1) the amount of mass accumulated
in agreement with the amount deduced from the
outburst lightcurve, (2) the low mass flow rate in the disk in agreement with 
the predictions from the ADAF based spectral fit, (3) agreement
between theoretical and observed recurrence time (58 years
assuming no outburst was missed in between). This
result assumed a 4$M_\odot$ black hole,
the favored value at this time. Meanwhile it became clear that the
black hole mass in A0620-00 is higher and that the distance is
smaller. (To get good agreement with the observations a slightly
higher viscosity would then be appropriate. Estimates
can be deduced from the earlier computations. In Fig.3 to account for
such uncertainty we also add peak luminosities for half the amount of 
matter accumulated evaluated before.)

\section{Theoretically determined peak luminosity in outburst}
We use the results for disk evolution during quiescence to evaluate 
the maximal mass flow rate in the hot disk in outburst. We assume that 
with the onset of the outburst the disk immediately assumes the
structure of the quasi-stationary accretion flow.
This is an approximation to the real situation where this process
takes a short time during which a small amount of gas already is accreted
before the outburst maximum is reached. 

\begin{figure}[b]
\includegraphics[width=8.cm]{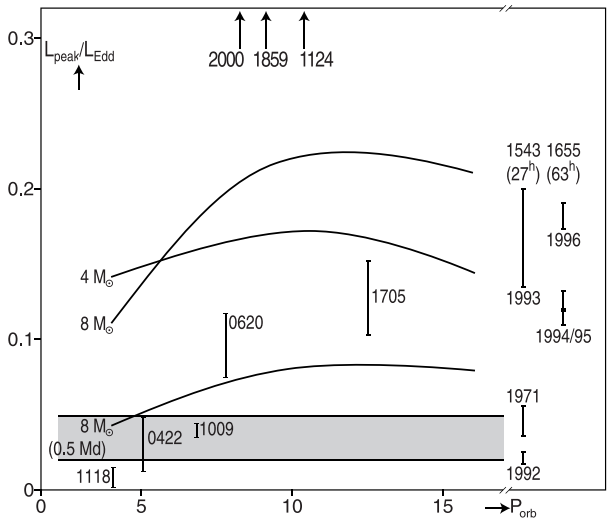}
\caption {Solid lines and dashed line: theoretically determined peak
luminosities in outburst as function of orbital period; gray area: 
range for critical luminosity for spectral state transition; 
circles: observational results for black hole X-ray transients, 
values larger than 0.3 are indicated by arrows in the uppermost part
of the figure.
}
\end{figure}

The mass flow rate 
in the hot disk was calculated according to the standard relation between
mass flow rate, mass in the disk, primary mass and disk size for a
viscosity parameter $\alpha = 0.2$ (Frank, King, and Raine 1985). As
expected, the mass flow rate is higher for disks with longer orbital
period. But for even larger disks (compare Fig. 3) the rates tend to 
level off due to the then longer diffusion time in the disk. For the 
correspondence between luminosity and mass accretion rate we used the 
formula $L=0.1 \dot M c^2$ ergs/s. This relation is appropriate
for the soft state and should hold for the transition. For the hard
state the luminosity might be somewhat smaller which could appear as
lower values in the observations. The
derived peak luminosities have to be compared with the critical value
for the spectral transition. Though as noted above our procedure
probably overestimates the absolute values of the peak luminosities
the curves in Fig. 3 can be used as an illustration of the trend:
low peak luminosities for systems with short orbital periods.
Therefore one would expect that those systems that remain in the hard state
would generally have short orbital periods.

\section{Comparison with observations}  
We now test whether the observed peak luminosities of systems
which show a hard to soft state transition lie above the critical
luminosity and those of systems which remain in the hard state in outburst
lie below the critical luminosity. This can only be performed for
systems where black hole mass and distance are known.
This reduces the comparison to a small number of
binaries (compare also Maccarone 2003). We list the data 
used here in Table 1.

\begin{table*}
\pagestyle{empty}
\setlength{\topmargin}{-2.5cm}
\setlength{\textwidth} {17cm}
\setlength{\oddsidemargin}{-2.0cm}
\setlength{\footskip}{-1.0cm}
\caption{ Peak luminosities \  of \  X-ray transients \ established as
black--hole sources}
\begin{center}
\begin{tabular}{lllllllll}
\hline
\hline
& & & & & & & &\\
Source & alternative & BH mass & companion & orbital & outburst &
distance & log$ L_{\rm{peak}}$ &
 \large{$\frac{L_{\rm{peak}}} {L_{\rm{Edd}}}$} \\
name & name & & star & period & year & & & \\
& & ($M_\odot$)& & (h)&  & (kpc)&  &\\
\hline
\\
XTE J1118+480 & KV UMa & 6.5-7.2 & K5/M0V &4.1 & 2000& 1.8 $\pm$ 0.5& 35.60 &0.0005\\
GRO J0422+32 & V518 Per & 3.97$\pm$0.95 & M1V & 5.1 & 1992 &
2.49$\pm$0.3 & 37.40 &0.049 \\ 
GRS1009-45& MM Vel&6.3-8.0&K7/M0V &6.8&1993& 5.0$\pm$ 1.3&37.99&0.106\\
A0620-00 &V616 Mon &8.7-12.9 &K4V &7.8 &1975 & 1.2$ \pm$
0.1&38.37& 0.169\\
GS2000+251& QZ Vul& 7.1-7.8&K3/K7V &8.3 &1988 & 2.7$\pm$ 0.7
&38.86&0.755\\
XTE J1859+226& V406 Vul&7.6-12:&- &9.2:& 1999& 11& 38.86& 0.574\\
GS1124-684&GU Mus &6.5-8.2 &K3/K5V &10.4 &1991 &5$\pm$ 1.3&38.78&0.636 \\
H1705-25&V2107 Oph&5.6-8.3 &K3/7V &12.5 &1977  & 8$\pm$2&38.58& 0.425 \\
4U1543-47&IL Lupi& 7.4-11.4&A2V&27.0 & 1971 &7.5$\pm0.5$&38.83&0.559 \\
& & & & & 1992&&38.48&0.249\\
GRO J1655-40&V1033 Sco &6.0-6.6 &F3/F5IV &62.9 &1994 &3.2$\pm0.2$ &37.86&0.089\\
& & & & & 1995&&37.77&0.073\\
& & & & & 1996&&38.02&0.129\\
\hline
\end{tabular}
\end{center}
\vspace {0.3cm}
\small{Note:
Black hole mass, spectral type of companion star, orbital period,
outburst year and distance as listed in McClintock \& Remillard
(2003). Peak luminosities in outburst $L_{\rm{peak}}$ taken from 
Chen et al. (1997, Table 8), values here determined with respect to 
the more recent black hole mass and distance estimates given above; Values for
XTE J1118+480 from Remillard et al. (2000), for GRO J0422+32 from
Gelino \& Harrison (2003), for XTE J1859+226 from Markwardt
(2001). Eddington luminosities according to the listed black hole
masses.} 
\end {table*}

\subsection{Observed peak luminosities in outburst}
We take the peak luminosities given by Chen, Shrader \& Livio (1997)
available at that time, for XTE J1118+480 the value from Remillard et 
al. (2000) and for XTE J1859+226 the value from Markwardt (2001). We did
not include the data for the 1983 outburst of 4U1543-47 because of the 
only very short time observation.
During the last years improved estimates for black hole masses and
distances have been obtained for
several systems, e.g. for A620-00 the black hole
mass accepted earlier was 4.9-10 $M_\odot$ (Chen et al. 1997), while now
McClintock and Remillard (2003) obtain 8.7-12.9 $M_\odot$ and the
distance estimate changed from 0.87 to 1.2 kpc. 
New optical and infrared photometry of GRO J0422+32 (Gelino \&
Harrison 2003) lead to a determination of the black hole mass of
3.97$\pm$0.95 $M_\odot$ with an adopted distance of 2.49$\pm$0.30 kpc.
We have not included in
our analysis sources with orbital periods much longer than the range
covered by our evolutionary results. Certainly e.g. GRS 1915+105 would
be an interesting example to study the spectral state transitions, but
for the larger accretion disks additional investigations would be
needed to understand the observations.
In Table 1 we give for all  sources considered here.
the values of the peak luminosity in outburst according to the new
distance estimates The uncertainty in distance (up to 25\%)
causes an uncertainty in the peak luminosity. In the Eddington
luminosity an even larger uncertainty arises from the uncertainty of the 
black hole mass (we used the arithmetic mean value). We plot the
observed peak luminosities in Fig. 3 but the just mentioned
uncertainties in $L_{\rm{peak}}/ L_{\rm{Edd}}$ should be kept in mind.

The observed peak luminosities differ surprisingly from source to
source. While some values lie in the same range as the 
theoretically predicted numbers the peak luminosity of the sources 
GS 2000-251, XTE J1859+226, GRS 1124-684 and H1705-25 is very high, up
to significant fractions of the Eddington luminosity.
From disk evolution for sources with similar disk size and black hole
mass one would expect a similar amount of matter accumulated. As
mentioned earlier a difference in the mass overflow
rate from the secondary star has little influence as long as we
consider transient sources with long recurrence time.
In these systems a high mass transfer rate is excluded, it would lead 
to a short outburst recurrence time. Such a large amount of gas
as is necessary for the observed high mass flow rate in outburst can only be
accumulated in the disk if the viscosity is particularly low, possibly
related to a low magnetic activity of the secondary star.

\subsection{Luminosities at spectral state transitions}
In our analysis we compare peak luminosity with the critical
luminosity for state transition. Theory predicts this value to be about
0.02 in Eddington luminosity (Meyer et al. 2000a). This is strongly
supported by the recent analysis of observational data by Maccarone 
(2003) who finds that the luminosities at soft to hard state transition
of 4 transient black hole X-ray sources, 2 persistent black hole
binaries and 3 neutron stars all lie in the range of 1-4\% of the
Eddington luminosity. To observe the hard to soft transition, usually
during the early outburst rise, is more difficult. The observations
seem to indicate that the luminosity at which that transition occurs is
higher than that of the transition back to the hard state. 
The existence of such a ``hysteresis effect'' was first pointed out by 
Miyamoto et al. (1995), indicated also by later observations
for a number of systems. The difference in luminosity lies around a
factor of five (for a recent discussion see e.g. Zdziarski et
al. 2004). In our Fig. 3 we take a range for the critical luminosity.
The ``hysteresis effect'' will be investigated in work in progress 
(Meyer-Hofmeister et al., 2004). 

\subsection{Sources which remain in the low/hard state during the outburst} 
Observations have shown that a few sources have spectra which
do not change to the soft/high state during an outburst, but
remain in the hard state. Brocksopp et al. (2001) list the following systems:
GS 1354-64, GRO J0422+32, GRS 1719 (1716) and XTE J1118+480. GS 1354-64 had
three (or four) outbursts during two of which the spectrum remained
hard. It is worth noting that these two also had significantly lower 
count rates indicating significantly lower peak luminosities.
Another source that remains in the hard state, GRS 1737-31, was mentioned by
McClintock \& Remillard (2003). In very recent work Brocksopp et
al. (2004) list a few more sources as possible hard state candidates.
Of the systems listed above only for GRO J0422+32 and XTE J1118+480 black hole 
masses are known. These systems are plotted in Fig. 3. Indeed the peak 
luminosities are very low and the orbital periods are short as expected.

\section*{Conclusions}

The peak luminosities reached during outburst discriminate between
systems in which, for high peak luminosity, a transition to the soft
spectral state occurs, and, those in which, for low peak luminosity, 
the spectrum remains hard.  For low peak luminosity, i.e. low mass flow
rate in the disk, coronal evaporation truncates the thin disk even in 
outburst, further inward an advection-dominated accretion flow occurs 
and the spectrum stays hard. For high peak luminosity instead the disk
cannot be truncated, the thin disk reaches inward to the last stable
orbit and the spectrum is a soft multi-color black body spectrum.
We have discussed the dependence on disk size and black hole mass.
We expect the X-ray novae with the shortest orbital periods
to eventually remain in the low/hard state throughout the outburst. 
This agrees with the observations for XTE J1118+480 and GRO J0422+32. 

 For a few sources, GS 2000+251, XTE J1859+226 and GRS 1124-684 the
observed peak luminosity is very high. An earlier investigation
(Meyer-Hofmeister \& Meyer 2000) had shown that, in systems with long
recurrence time, the mass overflow rate from the secondary star only
weakly influences the mass accumulated during quiescence. Therefore 
the large amount of matter accumulated in these disks cannot be
due to an enhanced mass overflow rate. We argue a peak luminosity of 
a significant fraction of the Eddington luminosity must be related to
a lower viscosity in the cool disk.
Since the origin of the viscosity in the cool disk possibly lies in the
magnetic fields reaching over from the companion star (Meyer \&
Meyer-Hofmeister 1999) this leads to the question whether the magnetic
fields of the secondary stars of these systems are less strong.

\end{document}